\documentclass[12pt,preprint]{aastex}
\slugcomment{}

\begin{document}

\title{Constraints on cosmological models from strong gravitational lensing systems}

\author{Shuo Cao$^1$, Yu Pan$^{1,2}$, Marek Biesiada$^3$, Wlodzimierz Godlowski$^4$ and Zong-Hong Zhu$^{1}$}

\affil{$^1$ Department of Astronomy, Beijing Normal University,
Beijing 100875, China; zhuzh@bnu.edu.cn \\
$^2$ College Mathematics and Physics, Chongqing Universe of Posts
and Telecommunications, Chongqing 400065, China \\
$^3$ Department of Astrophysics and Cosmology, Institute of Physics,
University of Silesia, Uniwersytecka 4, 40-007 Katowice, Poland\\
$^4$ Institute of Physics, Opole University, Oleska 48, 45-052
Opole, Poland}

\begin{abstract}
Strong lensing has developed into an important
astrophysical tool for probing both cosmology and galaxies (their
structure, formation, and evolution). Using the gravitational
lensing theory and cluster mass distribution model, we try to
collect a relatively complete observational data concerning the
Hubble constant independent ratio between two angular diameter
distances $D_{ds}/D_s$ from various large systematic gravitational
lens surveys and lensing by galaxy clusters combined with X-ray
observations, and check the possibility to use it in the future as
complementary to other cosmological probes. On one hand, strongly
gravitationally lensed quasar-galaxy systems create such a new
opportunity by combining stellar kinematics (central velocity
dispersion measurements) with lensing geometry (Einstein radius
determination from position of images). We apply such a method to a
combined gravitational lens data set including 70 data points from
Sloan Lens ACS (SLACS) and Lens Structure and Dynamics survey
(LSD). On the other hand, a new sample of 10 lensing galaxy clusters
with redshifts ranging from 0.1 to 0.6 carefully selected from
strong gravitational lensing systems with both X-ray satellite
observations and optical giant luminous arcs, is also used to
constrain three dark energy models ($\Lambda$CDM, constant $w$ and
CPL) under a flat universe assumption. For the full sample ($n=80$)
and the restricted sample ($n=46$) including 36 two-image lenses and
10 strong lensing arcs, we obtain relatively good fitting values of
basic cosmological parameters, which generally agree with the
results already known in the literature. This results encourages
further development of this method and its use on larger samples
obtained in the future.
\end{abstract}


\section{Introduction}

Pioneering observations of type Ia supernovae (SNe Ia)
\citep{Riess1998,Perlmutter1999} have demonstrated that our present
universe is passing through an accelerated phase of expansion
preceded by a period of deceleration. A new type of matter with
negative pressure known as dark energy, has come up to explain the
present phase of acceleration. The simplest candidate of dark
energy, the cosmological constant ($\Lambda$), is consistent with
various observations such as more precise supernova data
\citep{Riess2004,Davis2007,Kowalski2008}, the CMB observations
\citep{Spergel2007,Komatsu09}, the light elements abundance from Big
Bang Nucleosynthesis \citep{Burles2001}, the baryon acoustic
oscillations (BAO) detected in SDSS sky survey \citep{Eisenstein05},
radio galaxies \citep{Daly2009}, and gamma-ray bursts
\citep{Amati2008}. However, various other models were proposed as
candidates of dark energy, such as the typical dynamical scalar
field called quintessence \citep{Caldwell1998}, phantom corrections
\citep{Caldwell2002}, a joint quintom scenario \citep{Feng2005} or
Chaplygin gas \citep{Kamenshchik01,Zhu2004,Biesiada05,Zhang2006}, to
mention just a few out of a long list. On the other hand there are
still many other ways to understand the accelerating universe, such
as Modified Friedmann Equation \citep{Freese2002,Zhu2004b} and
Dvali-Gabadadze-Porrati(DGP) mechanism \citep{DGP2000}. But until
now none of these models was demonstrated superior over the other.
Besides, while updating the current estimates of cosmological model
parameters, one should try to use new probes. Strongly
gravitationally lensed systems belong to this category. They can
provide the information on two angular diameter distances, $D_{ds}$
and $D_s$. One is the distance to the source and the other is that
between the defector and the source. Since angular diameter distance
depends on cosmological geometry, we can use their ratios to
constrain cosmological models.

The discovery of strong gravitational lensing in Q0957+561
\citep{1979Natur} opened up an interesting possibility to use strong
lens systems in the study of cosmology and astrophysics. Up to now,
strong lensing has developed into an important astrophysical tool
for probing both cosmology \citep{Zhu00a,Zhu00b,Cha03, Cha04, Mit05,
Zhu08a, Zhu08b} and galaxies (their structure, formation, and
evolution) \citep{Zhu97,MS98,Jin00,Kee01,KW01,Ofek03,Treu06a}. Now
several hundreds of strong lens systems produced by massive galaxies
have been discovered, but only $\sim 90$ galactic-scale strong
lenses with known redshift of the lens and the source and measured
image separation can form well-defined samples useful for
statistical analyses. These well-defined strong lenses are
particularly useful not only for constraining the statistical
properties of galaxies such as stellar velocity dispersions or
galaxy evolution \citep{CM03,Ofek03}, but also for constraining
cosmological parameters such as the present-day matter density
$\Omega_m$, dark energy density $\Omega_x$ and its equation of state
$w$ \citep{Cha03,Mit05}. For example, the Cosmic Lens All-Sky
Survey (CLASS) statistical data, which consists of 8958 radio
sources out of which 13 sources are multiply imaged
\citep{Bro03,Cha03} was first extensively used by \citet{Cha03}, who
found $\Omega_m \approx 0.3$ assuming a flat cosmology and
non-evolving galaxy populations. \citet{Mit05} reused this CLASS
statistical sample based on the velocity dispersion function (VDF)
of early-type galaxies derived
from the SDSS Data Release 1 (DR1; \citet{Sto02}). \citet{Zhu08a} reanalyzed 10 CLASS
multiply-imaged sources whose image-splittings are known to be
caused by single early-type galaxies to check the validity of the
DGP model with radio-selected gravitational lensing statistics. More
recently, the distribution of gravitationally-lensed image
separations observed in the Cosmic Lens All-Sky Survey (CLASS), the
PMN-NVSS Extragalactic Lens Survey (PANELS), the Sloan Digital Sky
Survey (SDSS) and other surveys was used by \citet{Cao11a}, who found $w<-0.52$ assuming a flat cosmology and
adopting semi-analytical modeling of galaxy formation. The idea of
using strongly gravitationally lensed systems, in particular
measurements of their Einstein radii combined with spectroscopic
data, for measuring cosmological parameters including the cosmic
equation of state was discussed in \citet{Biesiada06} and also in a
more recent papers \citep{Grillo08,Biesiada10,Biesiada11}.

On the other hand, galaxy clusters, as the largest dynamical
structures in the universe, are also widely used both in cosmology
and astrophysics. Firstly, their mass distributions at different
redshifts can be described by the Press-Schechter function
\citep{Press1974}, which reflects the linear growth rate of density
perturbations and therefore can provide constraints on cosmological
parameters such as the matter and dark energy densities
\citep{Borgani1999}. Secondly, combining the Sunyaev-Zel'dovich
effect \citep{Sunyaev1972} with observations of clusters' X-ray
luminosity, one is able to measure or estimate the Hubble constant
and other cosmological parameters in given cosmological model
\citep{Reese2002,Schmidt2004,Jones2005,Bonamente2006,2004Zhu2}.
Relevant discussions on the corrections to the Sunyaev-Zeldovich
effect for galaxy clusters can be found in
\citet{Itoh98,Nozawa98,Nozawa06}. More importantly, giant arcs
generated by the galaxy cluster are perfect indicators of its
surface mass density, while the mass distribution of the cluster's
mass halo can be modelled from X-ray luminosity and temperature,
which may provide certain observable \citep{Sereno2002,Sereno2004}.
Recently, \citet{Yu10} collected a new sample with such
data from an online database BAX and various literature, which led
to some interesting results compared with those obtained by \citet{Sereno2004}.

In this paper, we try to collect a relatively complete observational
data concerning the Hubble constant independent ratio between two
angular diameter distances $D_{ds}/D_s$ from various large
systematic gravitational lens surveys and galaxy cluster data. This
paper is organized as follows. In Section~\ref{sec:method}, we
briefly describe the methodology for both strong gravitationally
lensed systems: galactic lenses and galaxy clusters. Then, in
Section~\ref{sec:data} we present the $D_{ds}/D_s$ data from various
large systematic gravitational lens surveys and lensing galaxy
clusters with X-ray observations and optical giant luminous arcs. We
further introduce three popular cosmological models tested in
Section~\ref{sec:model}. Finally, we show the results of
constraining cosmological parameters using MCMC method and conclude
in Section~\ref{sec:result}.

\section{The Method} \label{sec:method}
Gravitational lensing is one of the successful predictions of
General Relativity. Strong gravitational lensing occurs whenever the
source, the lens and the observer are so well aligned that the
observer-source direction lies inside the so-called Einstein ring of
the lens. \citet{Paczynski1981} tried to use lensing images as indicators to estimate cluster mass and
constrain cosmological constant.

In a cosmological context the source is usually a quasar with a
galaxy acting as the lens. Strong lensing reveals itself as multiple
images of the source, and the image separations in the system depend
on angular diameter distances to the lens and to the source, which
in turn are determined by background cosmology. Since the discovery
of the first gravitational lens the number of strongly lensed
systems increased to a hundred (in the CASTLES data base) and is
steadily increasing following new surveys like the Sloan Lens ACS
(SLACS) survey \citep{Newton11}. This opens a possibility to constraining the
cosmological model provided that we have good knowledge of the lens
model.

Now, the idea is that the formula for the Einstein radius in a SIS
lens (or its SIE equivalent),
\begin{equation}\label{ringeq}
\theta_E=4 \pi \frac{D_A(z,z_s)}{D_A(0,z_s)}
                 \frac{\sigma_{SIS}^2}{c^2},
\end{equation}
depends on the cosmological model through the ratio of (angular
diameter) distances between lens and source and between observer and
lens. Under flat Friedman-Walker metric, the angular diameter
distance reads
\begin{eqnarray}
\label{inted}
D_A(z;\textbf{p})=\frac{c}{H_{0} (1+z)}\int_{0}^{z} \frac{dz'}{E(z';p)}.
\end{eqnarray}
where $H_0$ is the Hubble constant and $E(z; \textbf{p})$ is a
dimensionless expansion rate dependent on redshift $z$ and
cosmological model parameters \textbf{p}. If the Einstein radius
$\theta_E$ from image astrometry and stellar velocity dispersion
$\sigma$ (or central velocity dispersion $\sigma_0$) from
spectroscopy can be determined, this method can be used to constrain
cosmological parameters. The advantage of this method is that it is
independent of the Hubble constant value and is not affected by dust
absorption or source evolutions. However, it depends on the
measurements of $\sigma_0$ and lens modelling (e.g. singular
isothermal sphere (SIS) or singular isothermal ellipsoid (SIE)
assumption). Hopefully, spectroscopic data for central parts of lens
galaxies became available from the Lens Structure and Dynamics (LSD)
survey and the more recent SLACS survey etc, which make it possible
to assess the central velocity dispersions $\sigma_0$
\citep{Treu06a,Treu06b,Grillo08}. Meanwhile, the SIS (or SIE) model
is still a useful assumption in gravitational lensing studies and
should be accurate enough as first-order approximation to the mean
properties of galaxies relevant to statistical lensing. For example,
\citet{Koopmans09} found that inside one effective radius massive
elliptical galaxies are kinematically indistinguishable from an
isothermal ellipsoid. In the previous works, such an isothermal mass
profile has also been widely used for analyses of statistical
lensing
\citep{Koc96,Kin97,Fassnacht98,RK05,Koopmans06,Koopmans09,Treu06a,Treu06b}.

However, let us note here that the velocity dispersion $\sigma_{SIS}$ of the mass distribution and
the observed stellar velocity dispersion $\sigma_{0}$ need not be the same.
\citet{White96} argued that there is a strong indication that dark matter halos are dynamically hotter
than the luminous stars based on X-ray observations, and dark matter must necessarily have
a greater velocity dispersion than the visible stars. In this paper, we adopt a free parameter $f_{E}$ that relates the velocity dispersion
$\sigma_{SIS}$ and the stellar velocity dispersion $\sigma_0$ \citep{Kochanek92,Ofek03}:
\begin{equation}
\sigma_{SIS}=f_{E}\sigma_{0}.
\label{f_E}
\end{equation}
To be more specific, we have kept $f_{E}$ as a free parameter, since it mimics the effects of:
(i) systematic errors in the rms difference between $\sigma_{0}$ (observed stellar velocity dispersion) and $\sigma_{SIS}$ (SIS model velocity dispersion);
(ii) the rms error caused by assuming the SIS model in order to translate the observed image separation into $\theta_E$, since the
observed image separation does not directly correspond to $\theta_E$;
(iii) softened isothermal sphere potentials which tend to decrease the
typical image separations \citep{Narayan96},
and could be represented by $f_{E}$ somewhat smaller than $1$.

\citet{Martel02} found that the presence of the background matter tends to increase
the image separations produced by lensing galaxies from ray-tracing simulations in CDM models, though this effect is small, of order of $20\%$ or less.
\citet{Christlein00} showed that richer environments of early type galaxies may have a higher ratio of dwarf to giant galaxies than the field.
However, \citet{Keeton00} showed that this effect nearly cancels the effect of the background matter,
making the distribution of image separations significantly independent of environment. They predicted that lenses in groups have a mean image separation
which is $\sim0.2''$ smaller than that of lenses in the field. Therefore, all these above
factors can possibly affect the images separation by
up to $\pm20\%$, which may be mimicked by introducing $(0.8)^{1/2}<f_{E}<(1.2)^{1/2}$ \citep{Ofek03}.

In the method we use, the cosmological model enters not through a
distance measure directly, but rather through a distance ratio
\begin{equation}\label{ratioth}
\mathcal{D}^{th}(z_d,z_s;\textbf{p})= \frac{D_{ds}}{D_{s}} =\frac{\int_{z_d}^{z_s}[dz'/E(z';\textbf{p})]}{\int_{0}^{z_s}[dz'/E(z';\textbf{p})]}
\end{equation}
and respective observable counterpart reads
\begin{equation}\label{ratio1}
\mathcal{D}^{obs}= \frac{c^2 \theta_E}{4\pi \sigma_{0}^2 f_E^2}
\end{equation}
with its corresponding uncertainty calculated through propagation equation concerning the errors both on the stellar velocity dispersion $\sigma_{0}$ and the Einstein radius $\theta_E$ ($\sim5\%$ error for the Einstein radius \citep{Grillo08}).

Another source of systematic errors in our method
comes from the fact that Einstein radius estimation from observed image positions
depends on the lens model (SIS or SIE or the other realistic mass distribution).
Moreover, the image separation could be affected by nearby masses (satellites, neighbor galaxies) or
the structures along the line of sight. This last issue will also be discussed in the last section. Here, let us
note that formally, at the level of Eq.(\ref{ratio1}) the $f_E$ factor does the double job accounting for
systematics associated both with $\sigma_0$ and $\theta_E$. Since the main goal of this paper is to constrain cosmological parameters, we firstly consider $f_E$ as a free parameter, obtain its best-fit value and probability distribution function $P(f_E)$, and then treat it as a "nuisance" parameter to determine constraints on the relevant cosmological parameters of interest. The procedure of marginalization is carried out following that of \citet{Allen08,Samushia08}, where $P(f_E)$ is normalized to one and is usually taken to be a Gaussian or a $\delta(f_E)$ function peaked at the best-fit value of $f_E^*$. We then integrate the likelihood function,
\begin{equation}
\label{Lp}
\mathcal{L}(\textbf{p})=\displaystyle{\int}\mathcal{L}(\textbf{p},f_E) P(f_E)df_E
\end{equation}
and determine the best-fit values and confidence level contours from $\mathcal{L}(\textbf{p})$.

Moreover, strong lensing by clusters with galaxies acting as sources
can produces giant arcs around galaxy clusters, which can also be
used to constrain clusters' projected mass and cosmological
parameters \citep{Lynds1986,Breimer1992,Sereno2004}. When a galaxy
cluster is relaxed enough, the hydrostatic isothermal spherical
symmetric $\beta$-model \citep{Cavaliere1976} can be used to
describe the intracluster medium(ICM) density profile:
$n_{e}(r)=n_{e0}\left( 1+r^{2}/r^{2}_{c}\right) ^{-3\beta_{X} /2}$,
where $n_{e0}$ is the central electron density, $\beta_{X}$ and
$r_{c}$ denote the slope and the core radius, respectively. Assuming
that whole gas volume is isothermal (with the temperature $T_{X}$),
the gravity of relaxed cluster and its gas pressure should balance
each other according to the hydrostatic equilibrium condition. With
the approximation of spherical symmetry, the cluster mass profile
can be given by $M(r)=\frac{3k_{B}T_{X}\beta_{X} }{G\mu
m_{p}}\frac{r^{3}}{r^{2}_{c}+r^{2}}$ , where $k_{B}$, $m_{p}$ and
$\mu=0.6$ are, respectively, the Boltzmann constant, the proton
mass, and the mean molecular weight \citep{Rosati2002}. A
theoretical surface density can be derived as
\begin{eqnarray}
\Sigma_{th}= \frac{3}{2G \mu m_{p}}
\frac{k_{B}T_{X}\beta_{X}}{\theta_{c}}\frac{1}{D_{d}}.
\end{eqnarray}
Combining this with the critical surface mass density for lensing
arcs \citep{Schneider1992}
\begin{eqnarray}
\Sigma_{obs}= \frac{c^{2}}{4\pi G} \frac{D_{s}}{D_{d} D_{ds}}\sqrt{
\frac{\theta_{t}^{2}}{\theta_{c}^{2}}+1},
\end{eqnarray}
a Hubble constant independent ratio can be obtained
\begin{eqnarray}\label{ratio2}
\mathcal{D}^{obs}=\frac{D_{ds}}{D_{s}}\Big|_{obs}=\frac{\,\mu m_{p}c^{2}}{6
\pi}\frac{1}{k_{B}T_{X}\beta_{X}}\sqrt{
\theta_{t}\!^{2}+\theta_{c}\!^{2} }.
\end{eqnarray}
The X-ray data fitting results may provide us the above mentioned
relevant parameters such as $T_{X}$, $\beta_{X}$, and $\theta_{c}$.
The position of tangential critical curve $\theta_{t}$ is usually
deemed to be equal to the observational arc position $\theta_{arc}$.
In this paper we assume that the deflecting angle has a slight
difference with the arc radius angle,
$\theta_{t}=\epsilon\theta_{arc}$, with the correction factor
$\epsilon=(1/\sqrt{1.2})\pm0.04$ \citep{Ono1999}. The complete set
of standard priors and allowances of the above parameters included
in Eq. [\ref{ratio2}] can be found in Table 1 of \citet{Yu10}. The observational
$\mathcal{D}^{obs}$ and its corresponding uncertainty are also calculated through Eq. [\ref{ratio2}].

We stress here that the observational distance ratio $\mathcal{D}$
has both advantage and disadvantage. The positive side is that the
Hubble constant $H_0$ gets cancelled, hence it does not introduce
any uncertainty to the results. The disadvantage is that the power
of estimating $\Omega_m$ is relatively poor \citep{Biesiada10}.
Therefore we only attempt to fit $\Omega_m$ in the case of a
$\Lambda$CDM model (where it is the only free parameter in flat
cosmology). Then for both cases above (Eq.~[\ref{ratio1}] and
Eq.~[\ref{ratio2}]), we can fit theoretical models to observational
data by minimizing the $\chi^{2}$ function
\begin{equation}\label{chi} \chi^2(\textbf{p})=
\sum_{i}\frac{(\mathcal{D}_i^{th}(\mathrm{\textbf{p}})-\mathcal{D}_{i}^{obs})^{2}}{\sigma
_{\mathcal{D},i}^{2}}.
\end{equation}
where the sum is over the sample and $\sigma _{\mathcal{D},i}^{2}$
denotes the variance of $\mathcal{D}_{i}^{obs}$.

\section{Sample used} \label{sec:data}
For the Einstein ring data, we first use a combined sample of 70
strong lensing systems with good spectroscopic measurements of
central dispersions from the SLACS and LSD surveys
\citep{Biesiada10,Newton11}. Original data concerning the sample
can be found in \citep{Koopmans02,Koopmans03,Treu04,Treu06a}(see
for details). In our sample of 70 lenses
some have 2 images and some have 4. There are some general arguments
in favor of SIS model, but strictly speaking SIS lens should have
only 2 images \citep{Biesiada10,Biesiada11}, so one can try to use
only 2 image systems out of the full sample. Therefore we selected a
subsample of $n=36$ lenses, which is summarized in Table \ref{list}
where the names of lenses in the restricted sample are given in
bold.

As for the strong lensing arcs, redshifts and temperatures of the
galaxy clusters are always searched out directly from online
databases, such as CDS (The Strasbourg astronomical Data Center) or
NED (NASA/IPAC Extragalactic Database). \citet{Yu10} have chosen a
new database established especially for X-ray galaxy clusters --
BAX, which provides detailed information including $\beta$ and
$\theta_{c}$. They also used the fitting results of Chandra, ROSAT, ASCA satellites
and VIMOS-IFU survey \citep{Ota2004,Bonamente2006,Covone05,Richard07}.
The final statistical sample satisfy the following well-defined selection criteria.
Firstly, the distance between the lens and the source should be always smaller
than that between the arc source and the observer, $D_{ds}/D_{s}<1$,
which rules out half of selected lensing arcs. Secondly, the arcs
whose positions are too far from characteristic radius
($\theta_{arc}>3 \theta_{c}$) should also be discarded \citep{Yu10}.
At last \citet{Yu10} obtained a sample of 10 giant arcs with all
necessary parameters listed in Table \ref{list}.

Now the observational $D_{ds}/D_s$ data containing 80 data points
for cosmological fitting are summarized in Table \ref{list}, with
errors calculated with error propagation equation. We also list a
restricted sample containing 46 data points, which consists 36
two-image lenses and 10 strong lensing arcs.

\begin{table*}[ht]
\caption{Values of $\mathcal{D}=D_{ds}/D_{s}$ from lensing galaxy clusters and combined SLACS+LSD lens samples. The two-image lenses are written in bold. } \label{list}
\begin{center} {\tiny 
\begin{tabular}{lcccccccccc}

\hline
System & z$_d$ & z$_s$ & $\mathcal{D}^{obs}$ & $\sigma_\mathcal{D}$ & ref \\
\hline
MS 0451.6-0305  & 0.550 & 2.91 & 0.785 & 0.087 & \citet{Bonamente2006,Yu10} \\
3C220.1 & 0.61 & 1.49 & 0.611 & 0.530 & \citet{Ota2004,Yu10}\\
CL0024.0 & 0.391 & 1.675 & 0.919 & 0.430 & \citet{Ota2004,Yu10} \\
Abell 2390 &  0.228  & 4.05 & 0.737 & 0.053  & \citet{Ota2004,Yu10} \\
Abell 2667 & 0.226 & 1.034 & 0.837 & 0.124 & \citet{Ota2004,Covone05,Yu10} \\
Abell 68 & 0.255 & 1.6 & 0.982 & 0.225 & \citet{Bonamente2006,Richard07,Yu10} \\
MS 1512.4 & 0.372 & 2.72 & 0.734 & 0.330 & \citet{Ota2004,Yu10} \\
MS 2137.3-2353 & 0.313 & 1.501 & 0.778 & 0.105 & \citet{Ota2004,Yu10} \\
MS 2053.7  &  0.583 & 3.146 & 0.968 & 0.209 & \citet{Ota2004,Bonamente2006} \\
PKS 0745-191 & 0.103 & 0.433 & 0.818 & 0.065 & \citet{Ota2004,Yu10} \\
\textbf{SDSS J0037-0942} & 0.1955 & 0.6322  & 0.6825 & 0.1026 & \citet{Biesiada10}\\
\textbf{SDSS J0216-0813} & 0.3317 & 0.5235  & 0.3632  & 0.0684 & \citet{Biesiada10}\\
\textbf{SDSS J0737+3216} & 0.3223 & 0.5812  & 0.3039  & 0.0458 & \citet{Biesiada10}\\
\textbf{SDSS J0912+0029} & 0.1642 & 0.324   & 0.5325 & 0.0789 & \citet{Biesiada10}\\
SDSS J0956+5100 & 0.2405 & 0.47   & 0.414 & 0.0628 & \citet{Biesiada10}\\
SDSS J0959+0410  & 0.126 & 0.5349  & 0.5599  & 0.1152 & \citet{Biesiada10}\\
\textbf{SDSS J1250+0523} & 0.2318 & 0.795  & 0.6179 & 0.0996 & \citet{Biesiada10}\\
SDSS J1330-0148 & 0.0808 & 0.7115  & 0.7762 & 0.1184 & \citet{Biesiada10}\\
SDSS J1402+6321 & 0.2046 & 0.4814  & 0.6575 & 0.1166 & \citet{Biesiada10}\\
SDSS J1420+6019 & 0.0629 & 0.5352  & 0.8593 & 0.1268 & \citet{Biesiada10}\\
SDSS J1627-0053 & 0.2076 & 0.5241  & 0.5078 & 0.0779 & \citet{Biesiada10}\\
\textbf{SDSS J1630+4520} & 0.2479 & 0.7933  & 0.8114 & 0.1347 & \citet{Biesiada10}\\
\textbf{SDSS J2300+0022} & 0.2285 & 0.4635  & 0.5531 & 0.0951 & \citet{Biesiada10}\\
\textbf{SDSS J2303+1422} & 0.1553 & 0.517  & 0.8651 & 0.1519 & \citet{Biesiada10}\\
SDSS J2321-0939 & 0.0819 & 0.5324  & 0.896 & 0.1312 & \citet{Biesiada10}\\
Q0047-2808 & 0.485 & 3.595  & 0.8872 & 0.1606 & \citet{Biesiada10}\\
\textbf{CFRS03-1077} & 0.938 & 2.941  & 0.6834 & 0.1377 & \citet{Biesiada10}\\
HST 14176 & 0.81 & 3.399  & 0.9757  &  0.1795 & \citet{Biesiada10}\\
\textbf{HST 15433} & 0.497 & 2.092  & 0.929 & 0.2067 & \citet{Biesiada10}\\
\textbf{MG 2016} & 1.004 & 3.263  & 0.5035 & 0.1234 & \citet{Biesiada10}\\
SDSS J0029-0055  & 0.227  & 0.9313    &   0.6356  &   0.1317  &\citet{Newton11}\\
\textbf{SDSS J0044+0113}  &   0.1196 &   0.1965    &   0.3877  &  0.0573  &\citet{Newton11}\\
SDSS J0109+1500  &   0.2939 &   0.5248    &   0.3803  &   0.0766  &\citet{Newton11}\\
SDSS J0252+0039  &   0.2803 &   0.9818    &   1.3426  &   0.2636  &\citet{Newton11}\\
\textbf{SDSS J0330-0020}  &   0.3507  &   1.0709   &   0.8498  &   0.2109  &\citet{Newton11}\\
SDSS J0405-0455  &   0.0753 &   0.8098    &   1.0851  &   0.1628  &\citet{Newton11}\\
SDSS J0728+3835  &   0.2058 &   0.6877    &   0.9477  &   0.1448  &\citet{Newton11}\\
SDSS J0822+2652  &   0.2414 & 0.5941    &   0.6056  &   0.1004  &\citet{Newton11}\\
SDSS J0841+3824  &   0.1159 &   0.6567    &   0.9671  &   0.143  &\citet{Newton11}\\
\textbf{SDSS J0935-0003}  &   0.3475 &   0.467     &   0.1926  &   0.0437  &\citet{Newton11}\\
SDSS J0936+0913  &   0.1897 &   0.588     &   0.6409  &   0.0953  &\citet{Newton11}\\
SDSS J0946+1006  &   0.2219 &   0.6085    &   0.6927  &   0.1452  &\citet{Newton11}\\
\textbf{SDSS J0955+0101}  &   0.1109 &   0.3159    &   0.8571  &   0.159  &\citet{Newton11}\\
\textbf{SDSS J0959+4416}  &   0.2369 &   0.5315    &   0.5599  &   0.1152  &\citet{Newton11}\\
SDSS J1016+3859  &   0.1679 &   0.4394    &   0.6204  &   0.0963  &\citet{Newton11}\\
SDSS J1020+1122  &   0.2822 &   0.553     &   0.524   &   0.0931  &\citet{Newton11}\\
SDSS J1023+4230  &   0.1912 &   0.696     &   0.836   &   0.1454  &\citet{Newton11}\\
SDSS J1029+0420  &   0.1045 &   0.6154    &   0.7952  &   0.1231  &\citet{Newton11}\\
SDSS J1032+5322  &   0.1334 &   0.329     &   0.4082  &  0.0618  &\citet{Newton11}\\
SDSS J1103+5322  &   0.1582 &   0.7353    &   0.9219  &   0.159  &\citet{Newton11}\\
SDSS J1106+5228  &   0.0955 &   0.4069    &   0.6222  &   0.0928  &\citet{Newton11}\\
\textbf{SDSS J1112+0826}  &   0.273 &   0.6295     &   0.5052  &   0.0885  &\citet{Newton11}\\
SDSS J1134+6027  &   0.1528 &   0.4742    &   0.6687  &   0.1005  &\citet{Newton11}\\
\textbf{SDSS J1142+1001}  &   0.2218 &   0.5039    &   0.6967  &   0.1735  &\citet{Newton11}\\
\textbf{SDSS J1143-0144}  &   0.106 &   0.4019     &   0.8061  &   0.1182  &\citet{Newton11}\\
SDSS J1153+4612  &   0.1797  &   0.8751   &   0.7138  &   0.1305  &\citet{Newton11}\\
\textbf{SDSS J1204+0358}  &   0.1644 &   0.6307    &   0.6381  &   0.1132  &\citet{Newton11}\\
\textbf{SDSS J1205+4910}  &   0.215 &   0.4808     &   0.5365  &   0.0803  &\citet{Newton11}\\
\textbf{SDSS J1213+6708}  &   0.1229  &   0.6402   &   0.5783  &   0.0883  &\citet{Newton11}\\
\textbf{SDSS J1218+0830}  &   0.135 &   0.7172     &   1.0498  &   0.158  &\citet{Newton11}\\
\textbf{SDSS J1403+0006}  &   0.1888 &   0.473     &   0.6352  &   0.1332  &\citet{Newton11}\\
SDSS J1416+5136  &   0.2987  &   0.8111   &   0.8259  &   0.2134  &\citet{Newton11}\\
SDSS J1430+4105  &   0.285 &   0.5753     &   0.509   &   0.1266  &\citet{Newton11}\\
\textbf{SDSS J1432+6317}  &   0.123 &   0.6643     &   1.1048  &   0.1662   &\citet{Newton11}\\
\textbf{SDSS J1436-0000}  &   0.2852 &   0.8049    &   0.775   &   0.1563  &\citet{Newton11}\\
\textbf{SDSS J1443+0304}  &   0.1338 &   0.4187    &   0.6439  &   0.1  &\citet{Newton11}\\
\textbf{SDSS J1451-0239}  &   0.1254 &   0.5203    &   0.7262  &   0.1275  &\citet{Newton11}\\
\textbf{SDSS J1525+3327}  &   0.3583 &   0.7173    &   0.6526  &   0.1611  &\citet{Newton11}\\
\textbf{SDSS J1531-0105 } &   0.1596 &   0.7439    &   0.7628  &   0.1147  &\citet{Newton11}\\
\textbf{SDSS J1538+5817}  &   0.1428 &   0.5312    &   0.972   &   0.172  &\citet{Newton11}\\
\textbf{SDSS J1621+3931}  &   0.2449  &   0.6021   &   0.8042  &   0.1765  &\citet{Newton11}\\
SDSS J1636+4707  &   0.2282 &   0.6745    &   0.7093  &   0.1276  &\citet{Newton11}\\
\textbf{SDSS J2238-0754}  &   0.1371  &   0.7126   &   1.1248  &   0.1812   &\citet{Newton11}\\
SDSS J2341+0000  &   0.186 &   0.807      &   1.1669  &   0.2049  &\citet{Newton11}\\
\textbf{Q0957+561}  &   0.36  &   1.41       &   1.3103  &   0.1474  &\citet{Newton11}\\
PG1115+080  &   0.31  &   1.72       &   0.7036  &   0.1604  &\citet{Newton11}\\
\textbf{MG1549+3047} &   0.11 &   1.17        &   0.5728  &   0.1194  &\citet{Newton11}\\
Q2237+030  &   0.04  &   1.169       &   0.6685  &   0.22  &\citet{Newton11}\\
\textbf{CY2201-3201} &   0.32 &   3.9     &   0.8526  &   0.305  &\citet{Newton11}\\
B1608+656   &   0.63 &   1.39        &   0.646   &   0.2154  &\citet{Newton11}\\

 \hline

\end{tabular}}\\

\end{center}
\end{table*}

\section{Cosmological models tested} \label{sec:model}
All cosmological models we will consider in this paper are currently
viable candidates to explain the observed acceleration. Given the
current status of cosmological observations, there is no strong
reason to go beyond the simple, standard cosmological model with
zero curvature and cosmological constant $\Lambda$ (except for the
conceptual problems arising when one attempts to reconcile its
observed value with some estimate derived from fundamental
arguments \citep{Weinberg}). However, it is still
interesting to investigate alternative models. And we hope that
future observations of more accurate $D_{ds}/D_s$ data could allow
to better discriminate various competing candidates. In the MCMC
simulations we assume for each class of models the best fit values
of parameters found in the present work, and vary them within their
2$\sigma$ uncertainties. We assume spatial flatness of the Universe
throughout the paper, since it is strongly supported by independent
and precise experiments e.g. a combined 5-yr Wilkinson Microwave
Anisotropy Probe (WMAP5), baryon acoustic oscillations (BAO) and
supernova data analysis gives $\Omega_{tot}=
1.0050^{+0.0060}_{-0.0061}$ \citep{Hinshaw09}. Moreover, the
$\Omega_m= 0.27$ prior is used except in the $\Lambda$CDM model
where the fit is attempted.

For comparison we also performed fits to the newly released Union2
SNe Ia data (n=557 supernovae) from the Supernova Cosmology project
covering a redshift range $0.015 \leq z \leq 1.4$ \citep{Amanullah}.
In the calculation of the likelihood from SNe Ia, we marginalize
over the nuisance parameter \citep{Pietro03}
\begin{equation}
\chi^2_{\rm SNe}=A-\frac{B^2}{C}+\ln\left(\frac{C}{2\pi}\right),
\end{equation}
where $A=\sum_i^{557}{(\mu^{\rm data}-\mu^{\rm
th})^2}/{\sigma^2_i}$, $B=\sum_i^{557}{(\mu^{\rm data}-\mu^{\rm
th})}/{\sigma^2_i}$, $C=\sum_i^{557}{1}/{\sigma^2_i}$, and the
distance modulus is $\mu=5 \log(d_L/\rm{Mpc})+25$, with the
$1\sigma$ uncertainty $\sigma_i$ from the observations of SNe Ia;
and the luminosity distance $d_L$ as a function of redshift $z$
\begin{equation}
d_L=(1+z)\int^{z}_0\frac{cdz'}{H_0 E(z';\textbf{p})}~.
\end{equation}

\subsection{The standard cosmological model ($\Lambda$CDM)}\label{lambda}

We start our analysis by first setting out the predictions for the
current standard cosmological model. In the simplest scenario, the
dark energy is simply a cosmological constant, $\Lambda$, i.e.\ a
component with constant equation of state $w=p/\rho=-1$. If flatness
of the FRW metric is assumed, the Hubble parameter according to the
Friedmann equation is
\begin{equation}
E^2(z;\textbf{p})=\Omega_m(1+z)^3 + \Omega_\Lambda,
\end{equation}
where $\Omega_m$ and $\Omega_\Lambda$ parameterize the density of
matter and cosmological constant, respectively. Moreover, in the
zero-curvature case ($\Omega=\Omega_m+\Omega_\Lambda=1$), this model
has only one independent parameter: $\textbf{p}=\Omega_{m}$.

\subsection{Dark energy with constant equation of state ($w$CDM)}\label{constantw}
Allowing for a deviation from the simple $w=-1$ case, the
accelerated expansion is obtained when $w<-1/3$. In a zero-curvature
universe, the Hubble parameter for this generic dark energy
component with density $\Omega_x$ then becomes
\begin{equation}
E^2(z;\textbf{p})= \Omega_m(1+z)^3 + \Omega_x(1+z)^{3(1+w)}.
\end{equation}
Obviously, when flatness and $\Omega_m= 0.27$ are assumed, it is a
one-parameter model with the model parameter: $\textbf{p}=\{w\}$.

\subsection{Dark energy with variable equation of state (CPL)}\label{CPL}
If the equation of state of dark energy is allowed to vary with
time, one has to choose a suitable functional form for $w(z)$, which
in general involves certain parametrization. Now, we consider the
commonly used CPL model \citep{Chevalier01,Linder03}, in which the
equation of state of dark energy is parameterized as
$w(z)=w_0+w_a\frac{z}{1+z}$, where $w_0$ and $w_a$ are constants.
The corresponding $E(z)$ can be expressed as
\begin{equation} \label{cpl2}
E^2(z;\textbf{p})=\Omega_{m}(1+z)^3+(1-\Omega_{m})(1+z)^{3(1+w_0+w_a)}\exp\left(-\frac{3w_a
z}{1+z}\right).
\end{equation}
There are two independent model parameters in this model:
$\textbf{p}=\{w_0,~w_a\}$.

\begin{table}
\caption{\label{result} Fits to different cosmological models from
80 full $D_{ds}/D_{s}$ data and 46 restricted $D_{ds}/D_{s}$ data.
Fixed value of $\Omega_m=0.27$ is assumed except $\Lambda$CDM.}
\begin{center}
\begin{tabular}{clllllll}
\hline\hline
Cosmological model         &  Best-fitting parameters $(n=80)$ &    Best-fitting parameters $(n=46)$\\
\hline
$\Lambda$CDM       &    $\Omega_m=0.20^{+0.07}_{-0.07}$ &   $\Omega_m=0.26^{+0.11}_{-0.10}$ \\
  \hline
$w$CDM   &   $w=-1.02^{+0.26}_{-0.26}$    &      $w=-1.15^{+0.34}_{-0.35}$\\
  \hline
CPL      &   $w_0=0.60\pm1.76$    &   $w_0=-0.24\pm2.42$    \\
          &   $w_a=-7.37\pm8.05$  &  $w_a=-6.35\pm9.75$  \\

\hline
\end{tabular}
\end{center}
\end{table}

\begin{figure}
\begin{center}
\includegraphics[angle=0,width=70mm]{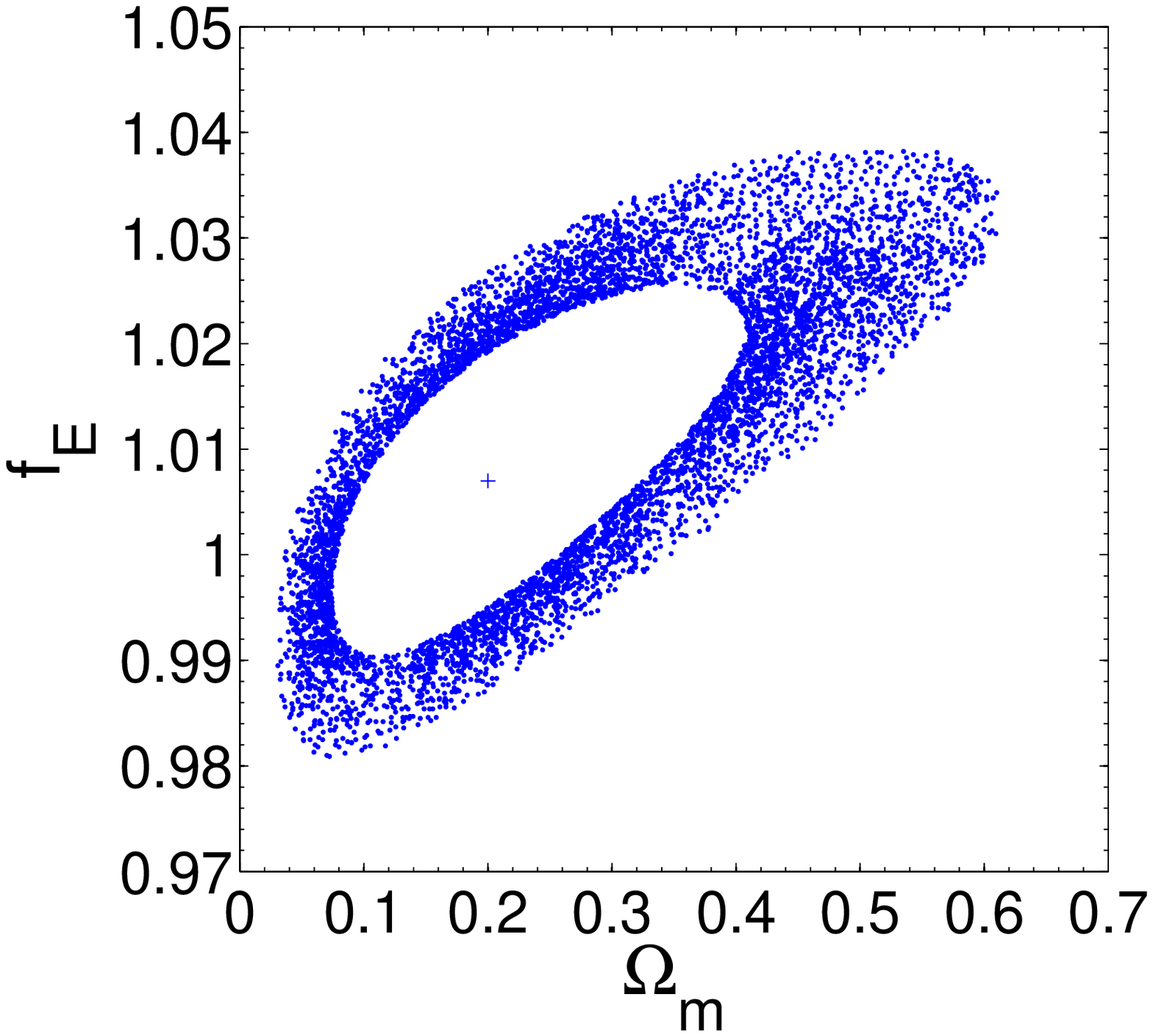}
\includegraphics[angle=0,width=70mm]{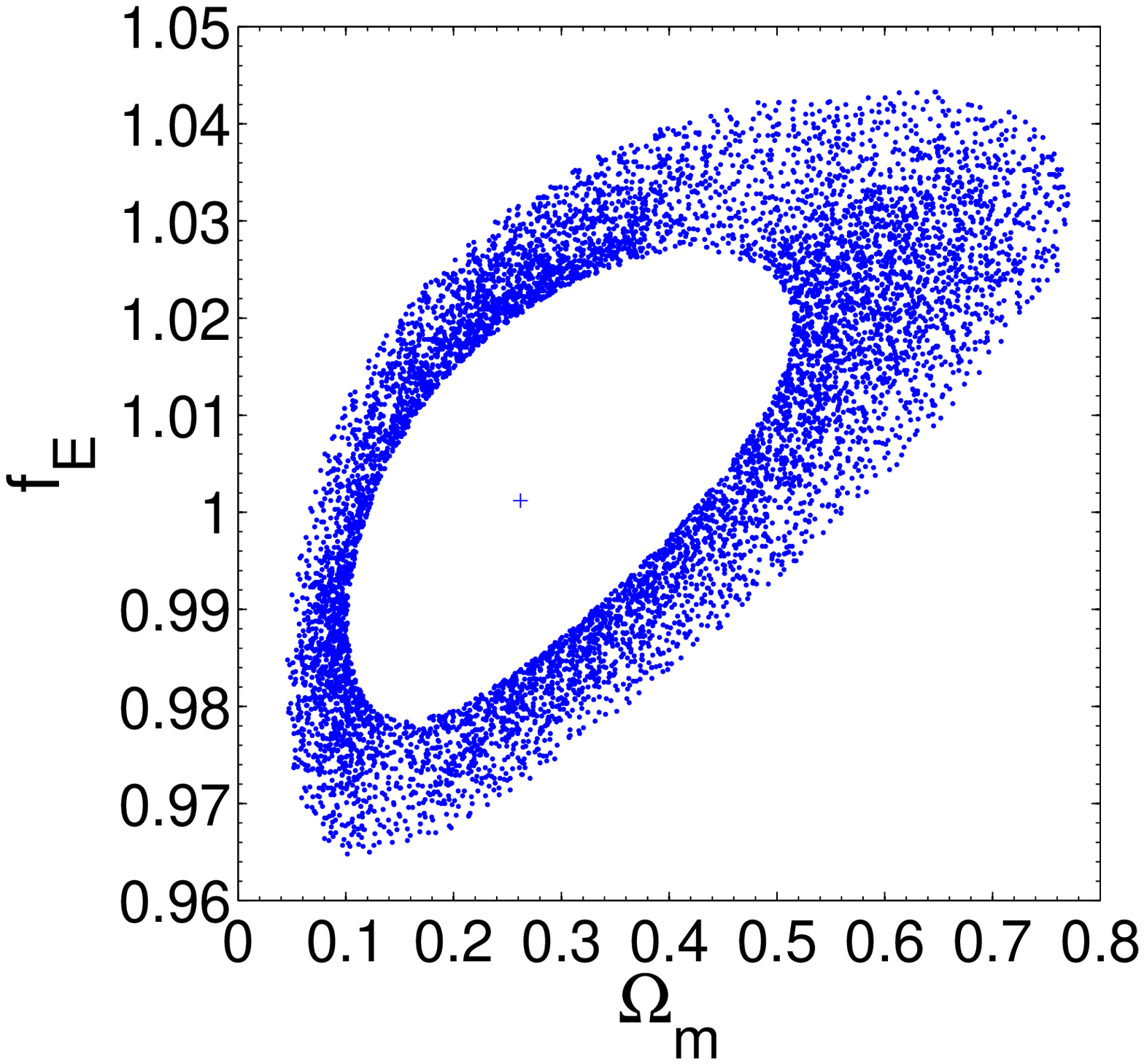}
\end{center}
\caption{ The 68.3 and 95.8 \% confidence regions
for $\Lambda$CDM model in the ($\Omega_m$,$f_E$) plane obtained from the full $n=70$ and the restricted $n=36$ two-image galaxy lenses. The crosses represent the best-fit points.
\label{L1}}
\end{figure}

\begin{figure}
\begin{center}
\includegraphics[angle=0,width=70mm]{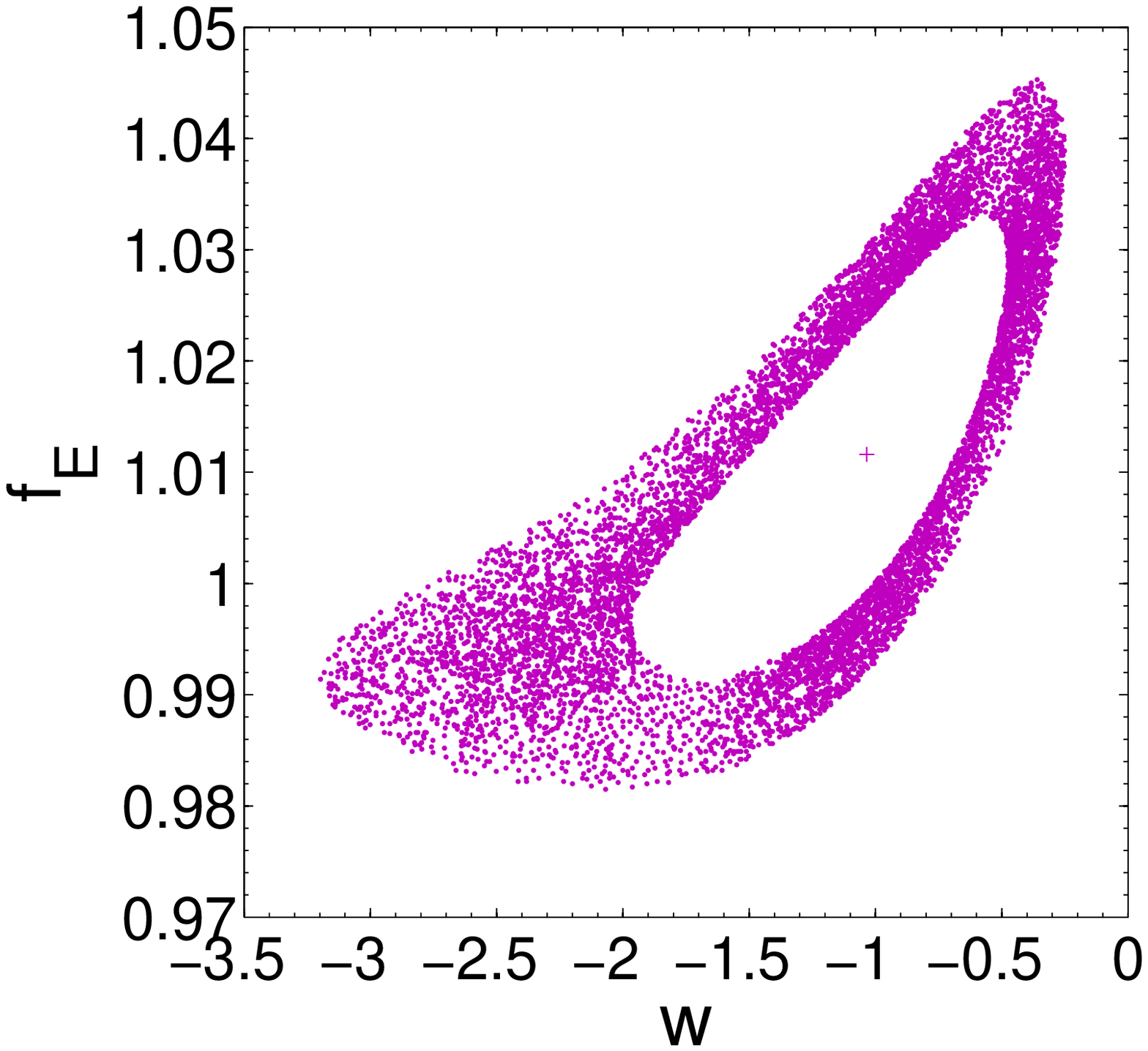}
\includegraphics[angle=0,width=70mm]{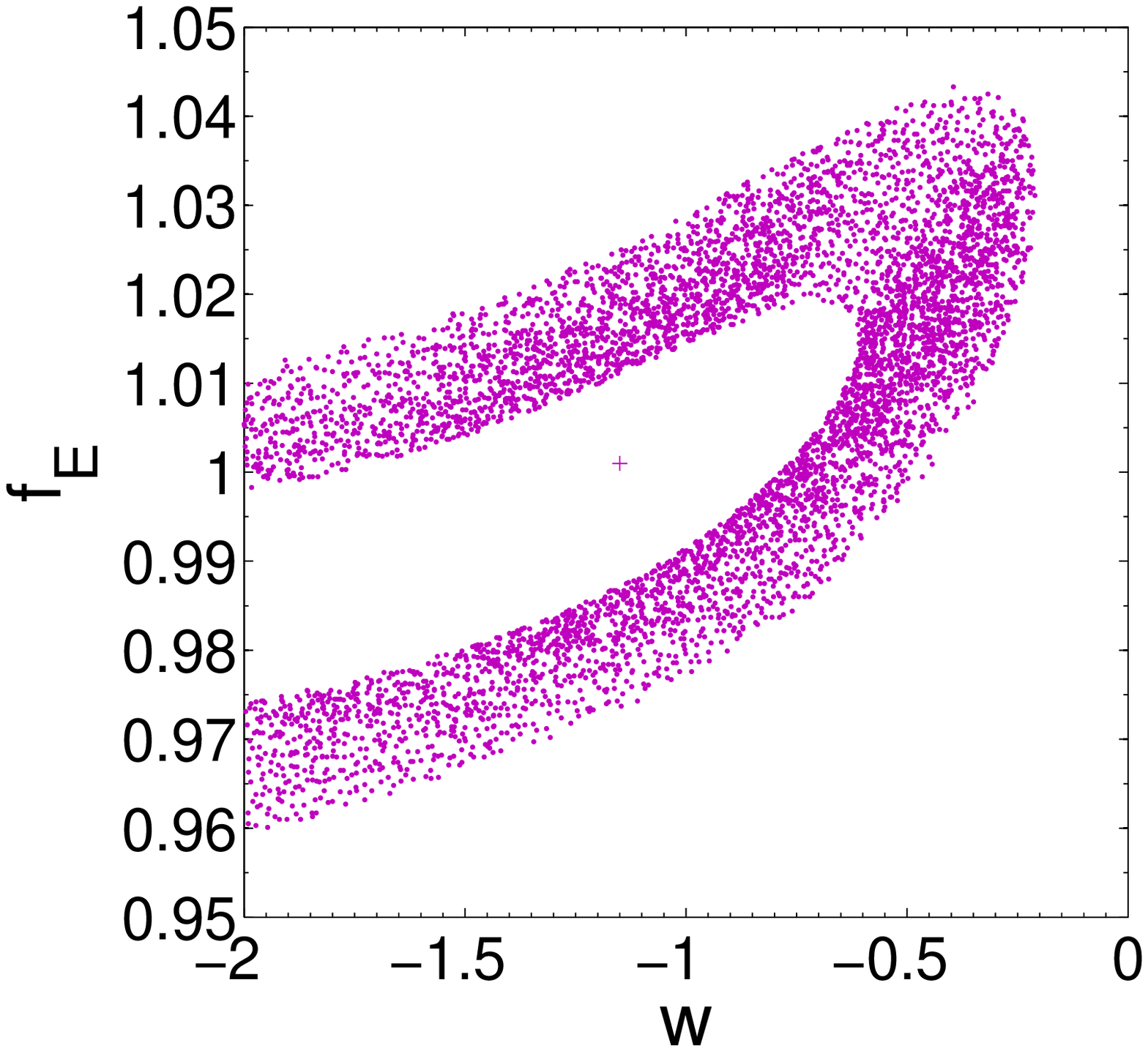}
\end{center}
\caption{The 68.3 and 95.8 \% confidence regions
for $w$CDM model in the (w,$f_E$) plane obtained from the full $n=70$ and the restricted $n=36$ two-image galaxy lenses. The crosses represent the best-fit points.
\label{w1}}
\end{figure}

\begin{figure}
\begin{center}
\includegraphics[angle=0,width=70mm]{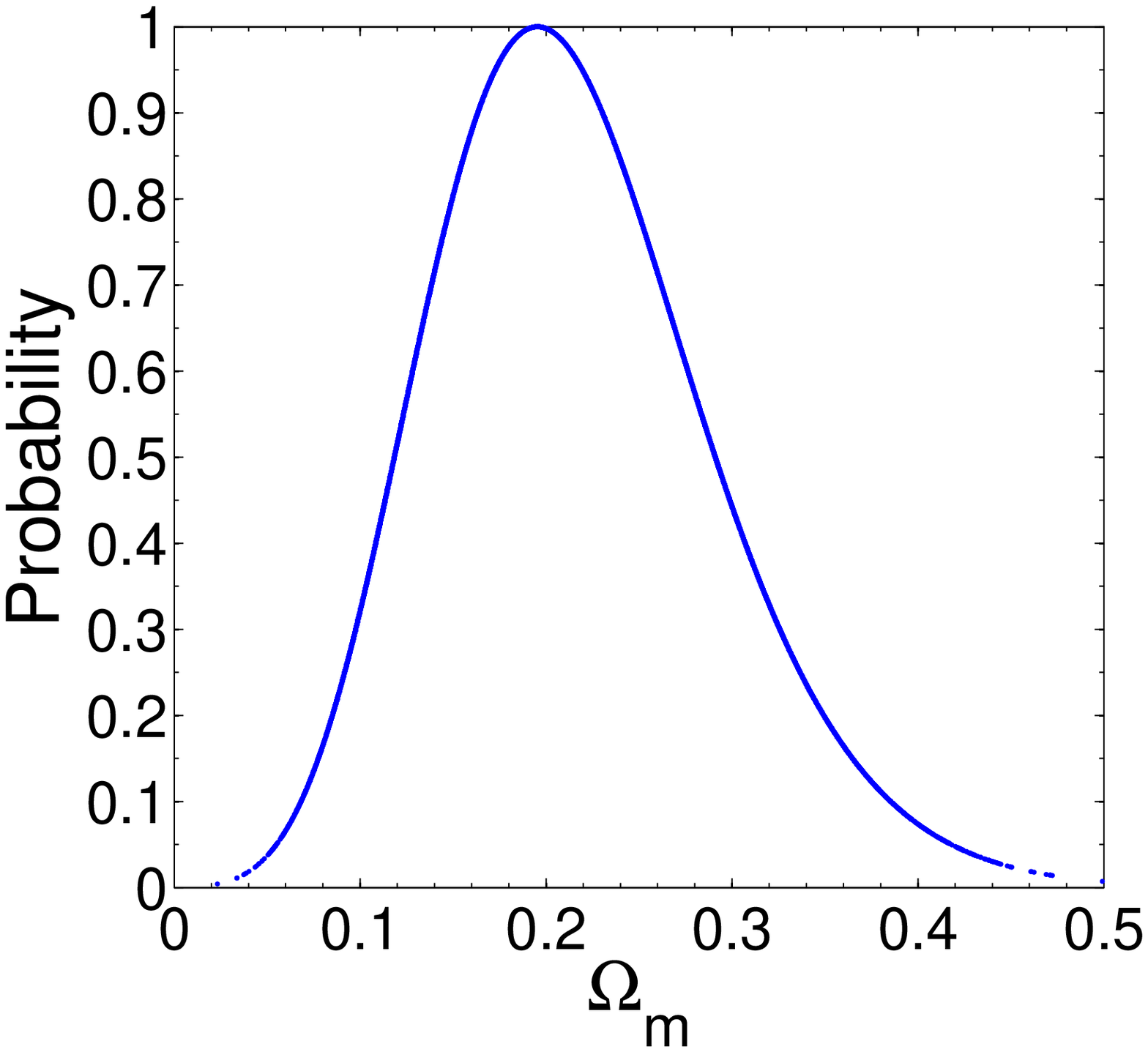}
\includegraphics[angle=0,width=70mm]{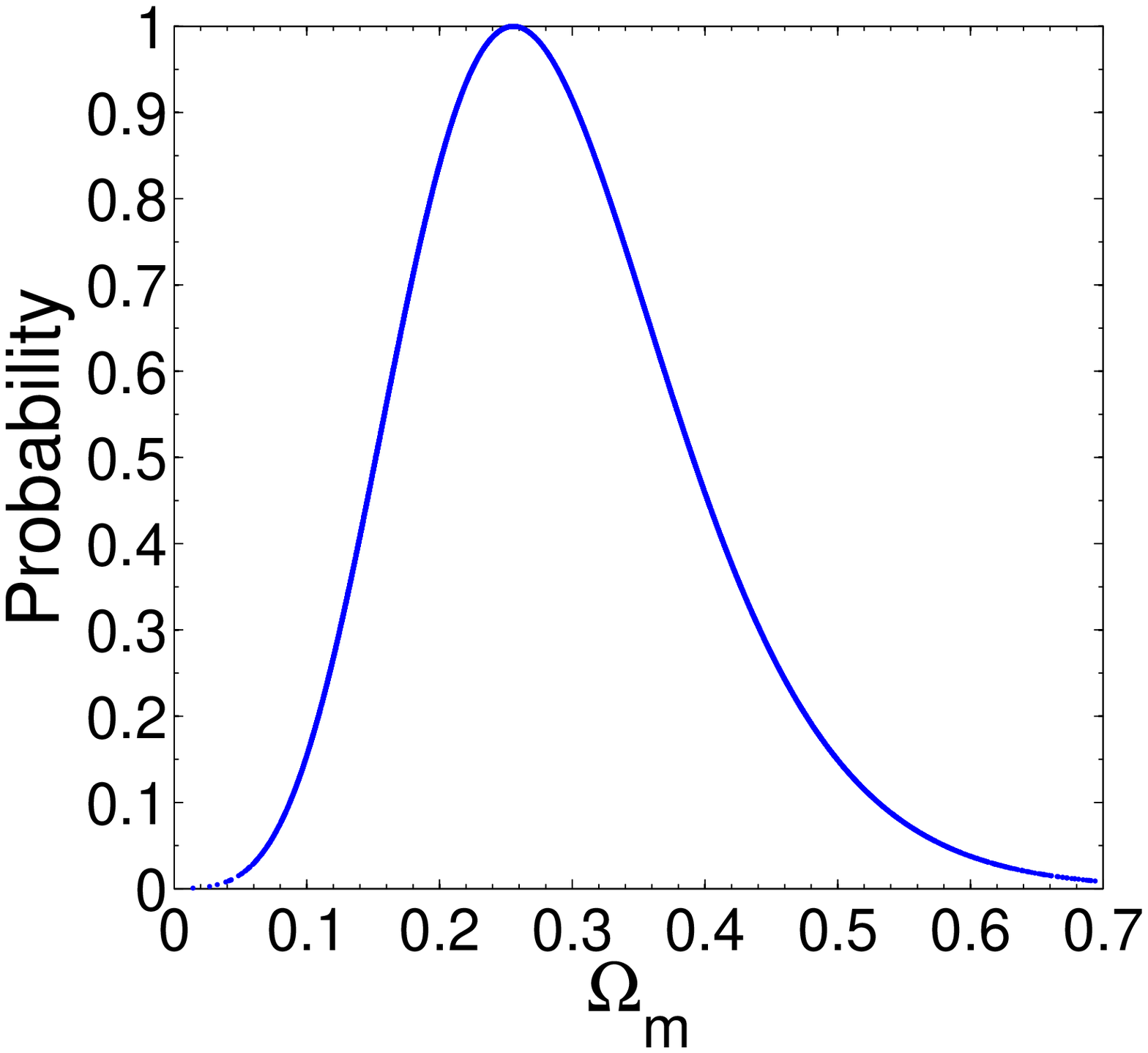}
\end{center}
\caption{The marginalized constraint on $\Omega_m$ of $\Lambda$CDM model from 80 full $D_{ds}/D_{s}$ data and 46 restricted $D_{ds}/D_{s}$ data.
\label{L}}
\end{figure}

\begin{figure}
\begin{center}
\includegraphics[angle=0,width=70mm]{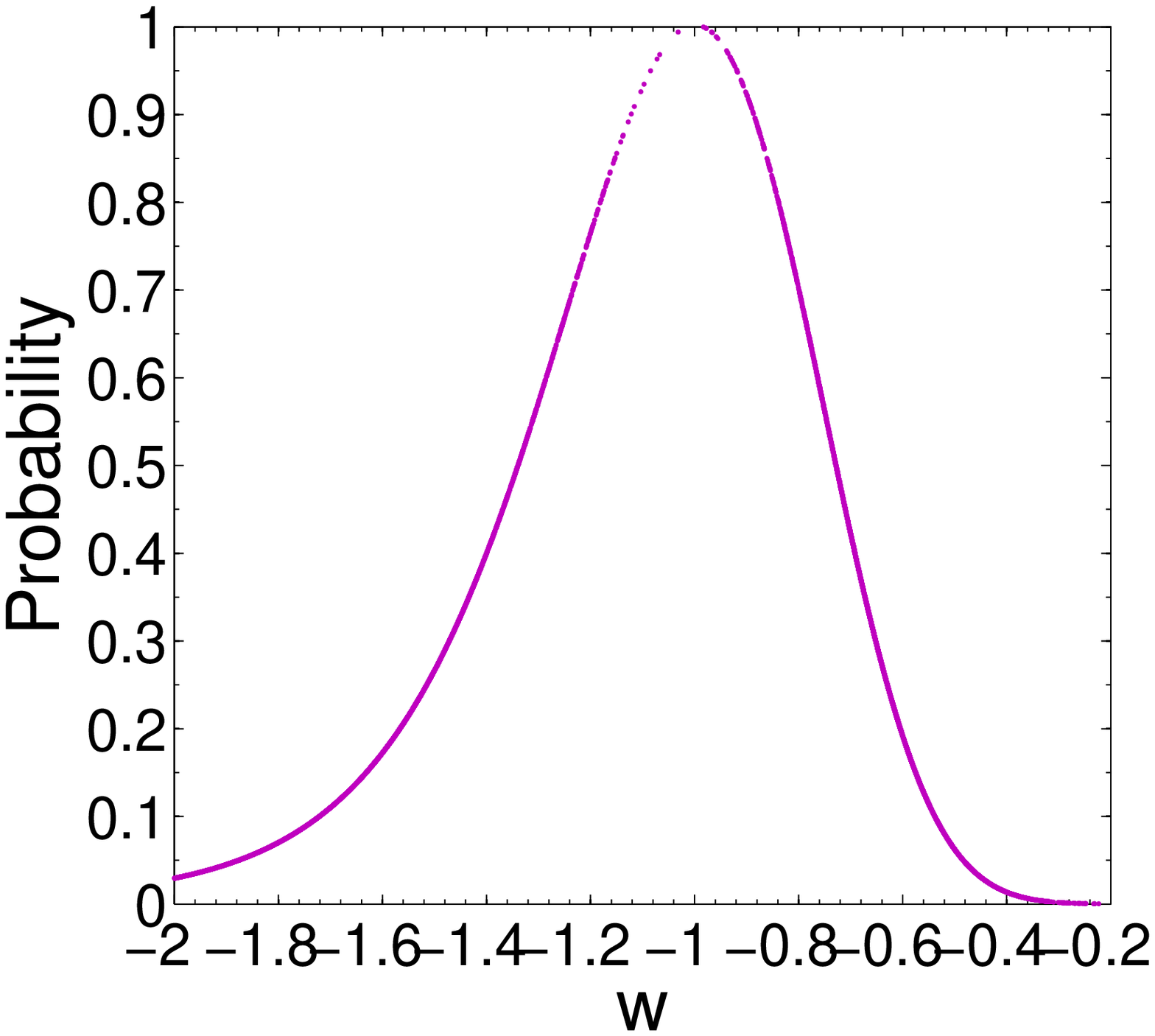}
\includegraphics[angle=0,width=70mm]{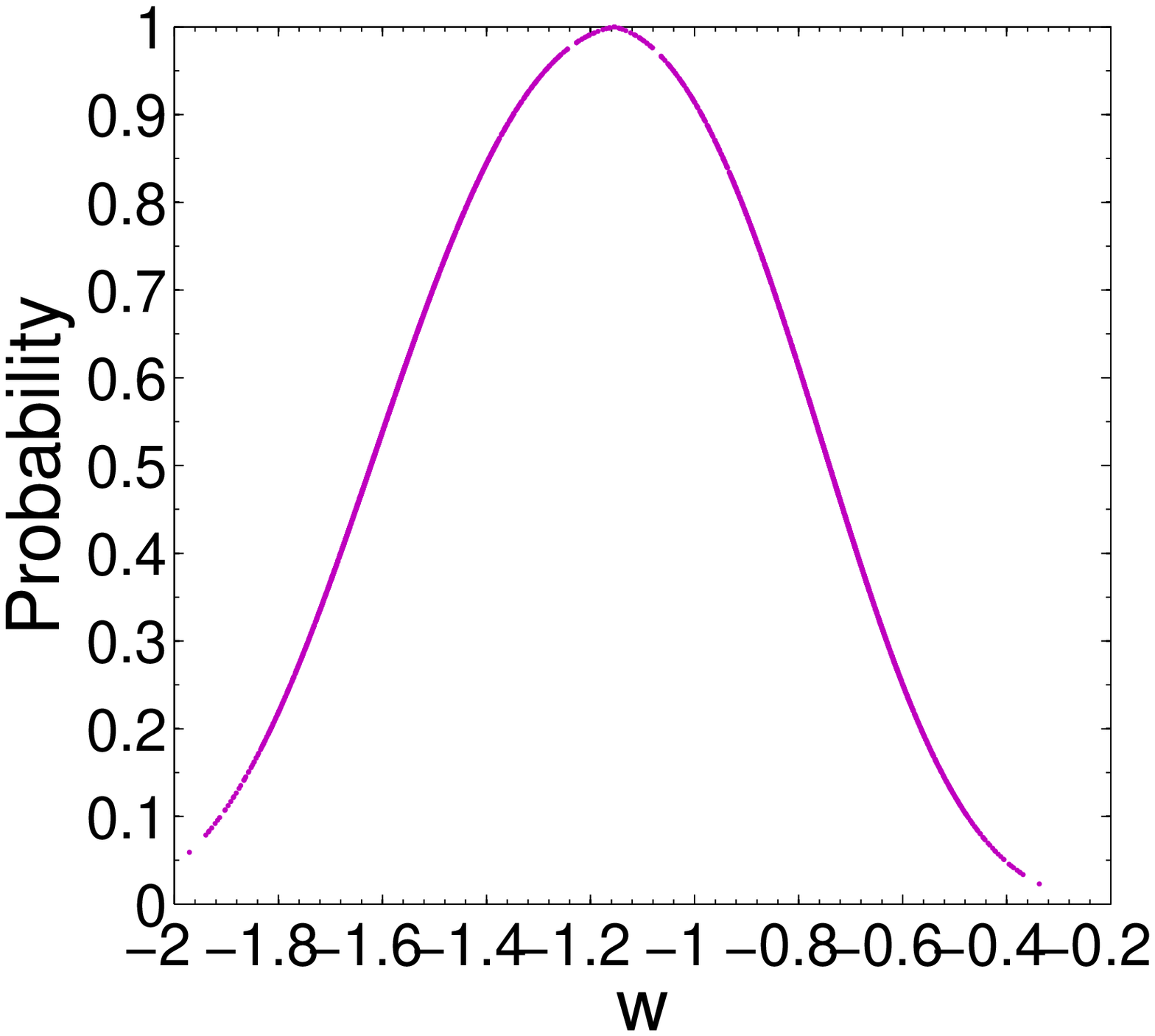}
\end{center}
\caption{The marginalized constraint on w of $w$CDM model from 80 full $D_{ds}/D_{s}$ data and 46 restricted $D_{ds}/D_{s}$ data.
\label{w}}
\end{figure}

\begin{figure}
\begin{center}
\includegraphics[angle=0,width=70mm]{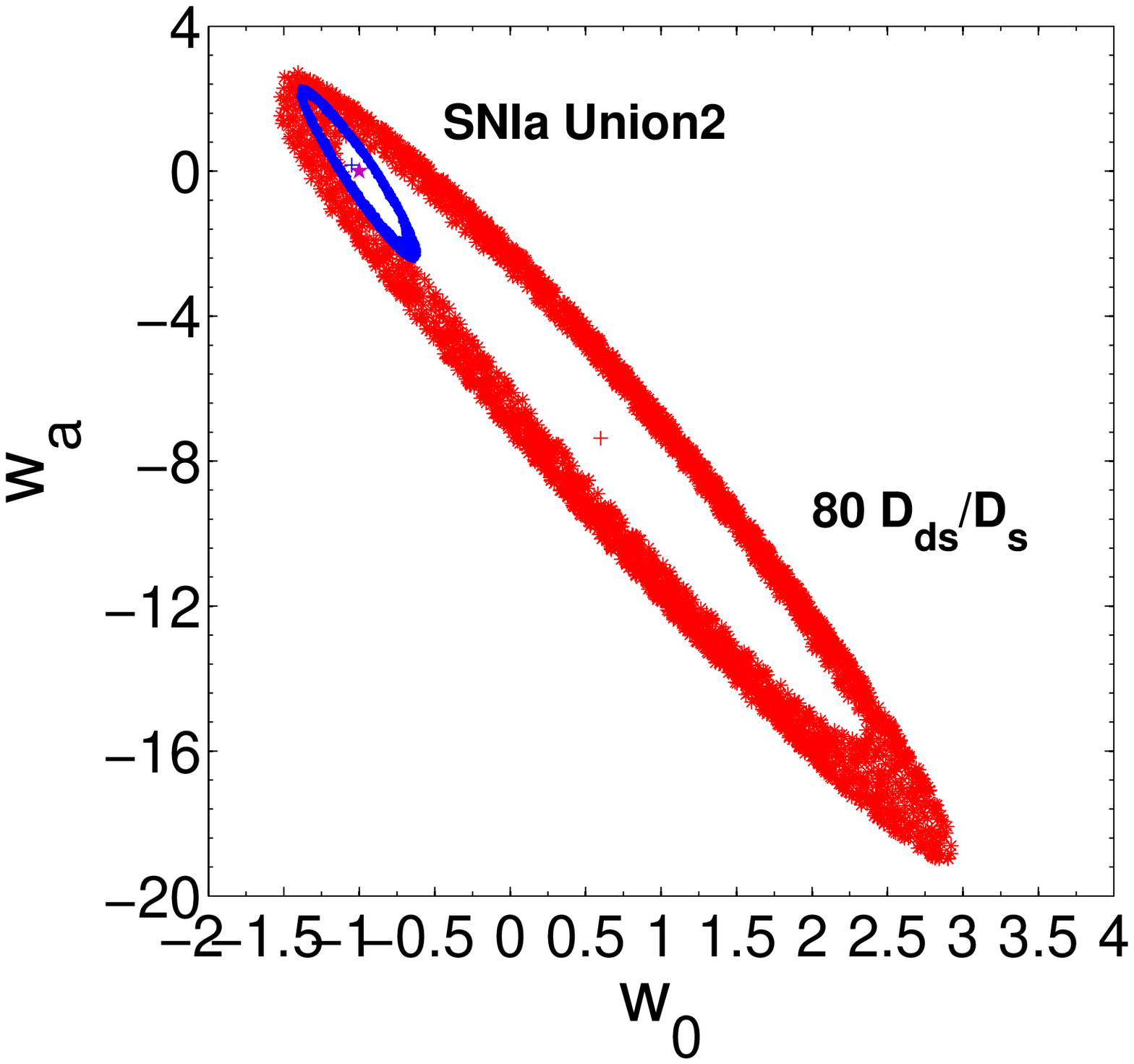}
\includegraphics[angle=0,width=70mm]{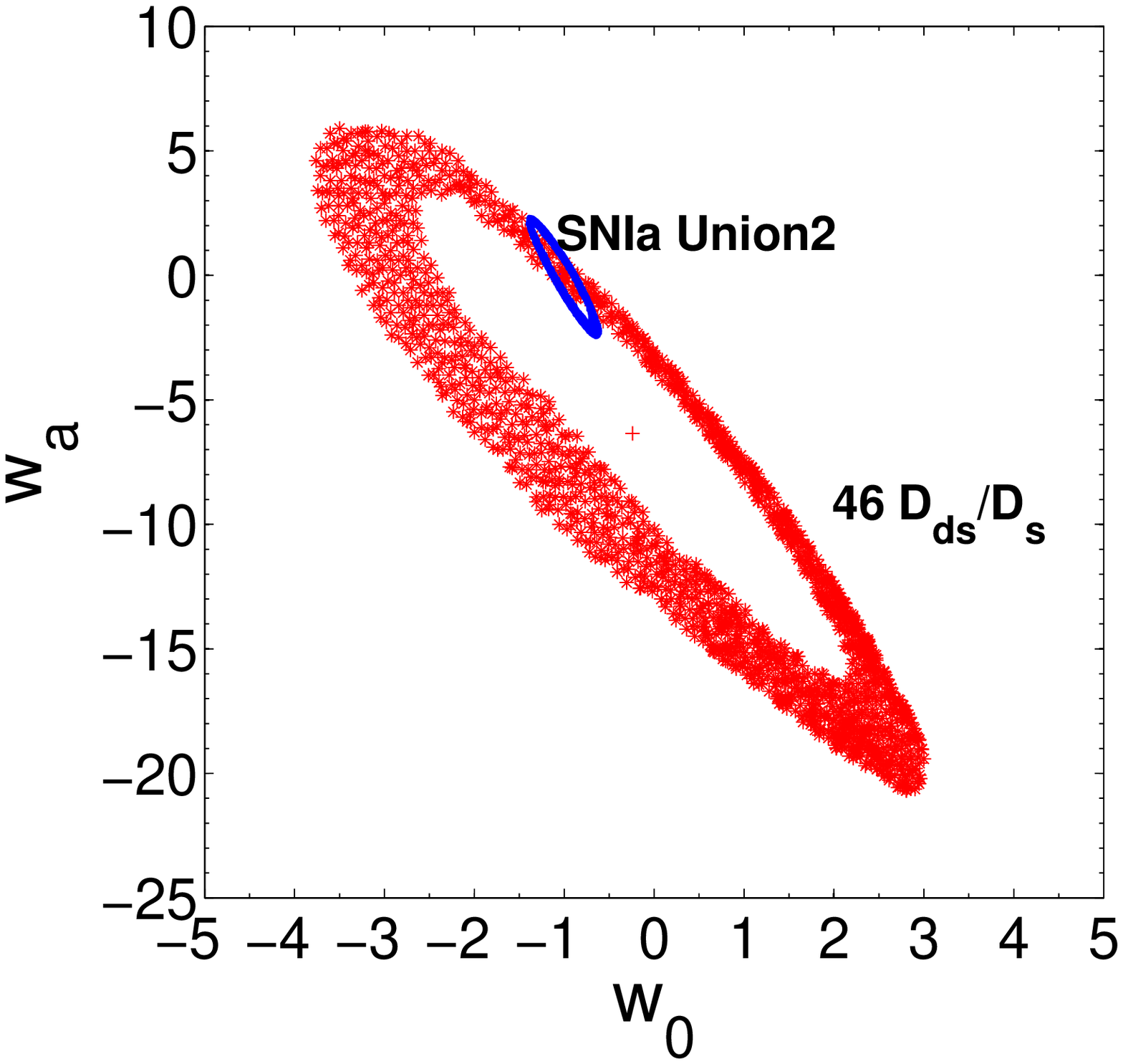}
\end{center}
\caption{ The 68.3 and 95.8 \% confidence regions
for CPL parametrization in the ($w_0$,$w_a$) plane obtained from 80 full $D_{ds}/D_{s}$ data, 46 restricted $D_{ds}/D_{s}$ data, and 557 Union2 SNe Ia
data. The crosses represent the best-fit points and a star corresponding to $\Lambda$CDM model is also added for reference.
\label{CPL}}
\end{figure}

\section{Results and conclusions} \label{sec:result}

In the first case, we consider $f_E$ as a free parameter and show the constraint results with the full $n=70$ and the restricted $n=36$ two-image galaxy lenses in Fig.~\ref{L1} and Fig.~\ref{w1}. In order to derive the probability distribution function for the cosmological parameters of interest, we marginalize $f_E$ through Eq. [\ref{Lp}] and perform fits of different cosmological scenarios on the full
$n=80$ sample as well as the restricted $n=46$ sample with the
results displayed in Table~\ref{result}.

For the full $n=80$ sample (containing 70 galaxy lenses and 10 strong lensing arcs), first, in $\Lambda$CDM model where $\Omega_m$ is the only free
parameter we were able to make a reliable fit on the samples
considered. This result is a considerable improvement over
\citet{Biesiada10}, where the authors failed to constrain $\Omega_m$
with their sample of twenty Einstein rings. Let us compare our results with
previously known ones. The current best
fit value from cosmological observations is: $\Omega_\Lambda=0.73\pm
0.04$ in the flat case \citep{Davis2007}. Moreover,
\citet{Komatsu09} gave the best-fit parameter $\Omega_{m}=0.274$ for
the flat $\Lambda$CDM model from the WMAP5 results with the BAO and
SNe Ia Union data. We find that our value of $\Omega_m$ (see Table 2) obtained from the
$D_{ds}/D_s$ data is consistent with the previous works at $1\sigma$.
Secondly, the best fit for the $w$CDM parameter agrees with that inferred from SNe Ia or WMAP5, and the $\Lambda$CDM model
($w=-1$) still falls within the 1$\sigma$ interval from the
$D_{ds}/D_s$ sample. Hence the agreement is quite good. Thirdly,
concerning the evolving equation of state in the CPL
parametrization, confidence regions in the ($w_0$,$w_a$) plane are
shown in Fig.~\ref{CPL}. One can see that fits for $w_0$ and $w_a$
are greatly improved as compared with those of \citet{Biesiada10}. The
values inferred are also in agreement with the WMAP5 results
presented in \citet{Hinshaw09} including combined WMAP5, BAO and SNe
Ia analysis. Moreover, it can be seen that the concordance model
($\Lambda$CDM) is still included at 1$\sigma$ level for the
$D_{ds}/D_s$ data applied here. For comparison we also plot the
likelihood contours with the Union2 SNe Ia compilation
\citep{Amanullah}. One can see that the $w$ coefficients obtained
from the $D_{ds}/D_s$ sample agrees with the respective values
derived from supernovae data (almost the whole 2$\sigma$ confidence
interval for $w$ from the Union2 data set lies within the 2$\sigma$
CI from the $D_{ds}/D_s$ data), which demonstrates the compatibility
between the SNe Ia and $D_{ds}/D_s$ data. This is also a great improvement over
 \citet{Biesiada10}, where SNe Ia results and
strong lensing results were found marginally inconsistent at 2$\sigma$.

Working on the restricted $n=46$ sample (containing 36 two-image
lenses and 10 strong lensing arcs), despite the sample size has
decreased dramatically, we find that fits on $\Omega_m$ in $\Lambda$CDM model are
consistent with the standard knowledge (see Fig.~\ref{L}) and the
best fit for the $w$ parameter in quintessence scenario is higher
than inferred from SNe Ia or WMAP5 (see Fig.~\ref{w}). Moreover,
for the fits on $w_0$ and $w_a$ in CPL parametrization, even though
confidence regions get larger in Fig.~\ref{CPL}, the result also turns
out to agree with SNe Ia fits. One should also note, that a systematic shift downwards in the ($w_0$,$w_a$) plane
persists. Such a shift in best-fitting parameters inferred from
supernovae (standard candles, sensitive to luminosity distance) and
BAO (standard rulers, sensitive to angular diameter distance) has
already been noticed and discussed in \citet{Linder08,Biesiada10}.
Our result suggests the need for taking a closer look at the
compatibility of results derived by using angular diameter distances
and luminosity distances, respectively. Recent discussions on the
ideas of testing the Etherington reciprocity relation between these
two distances can be found in \citet{Bassett04,Uzan04,HLB10,Cao11b,Biesiada11b}.

In conclusion our results demonstrate that the method extensively
investigated in \citet{Biesiada06,Grillo08,Biesiada10,Yu10} on
simulated and observational data can practically be used to
constrain cosmological models. Moreover, good quality measurements
of the relevant observational qualities such as the velocity
dispersion and Einstein radius turn out to be crucial. Finally, four
important effects, neglected here, should be mentioned. One is that
both the Einstein rings and X-ray observations of our new lensing
sample come from different surveys or satellites (SLACS, LSD and
SBAS and Chandra, ROSAT and ASCA, respectively), the differences in
detectors and observing strategies may cause systematical errors
which are hard to estimate. The second is that the observed image separation
is affected by secondary lenses (satellites, nearby galaxies, groups, etc) in many cases.
In this case, those lenses should not be used or the true $\theta_E$
corresponding to $\sigma_0$ should be estimated through realistic modelling.
However, most of our samples come from the SLACS survey where the role of environment
has been assessed in \citet{Treu09}. Namely, it was found that for SLACS lenses the typical contribution from external mass
distribution is no more than a few percent. The third important effect is, that the statistical procedure for cluster lenses relies on many simplifying assumptions. The realistic errors should be estimated by more realistic model of galaxy clusters besides the hydrostatic isothermal spherical symmetric $\beta$-model.The last one is the influence of line-of-sight mass contamination, with the significant effect of the
large-scale structure on strong lensing \citep{Bar-Kana96,Keeton97}.
More recent results on this issue can be found in \citet{Dalal05,Momcheva06}. In this paper, large scale structure effects which change the typical separation
between images are also included in the parameter $f_E$ - an increase of an arbitrary order $f_{E}^{1/2}$
in the velocity dispersion is equivalent to
an increase of $f_{E}$ in the typical separation $\theta$ (i.e., $\theta\propto\sigma^{2}$) \citep{Martel02}. In order to be complete with the discussion of possible errors one should also notice that the redshifts
$z_s$ and $z_l$ are also known with some accuracy $\delta z_s$ and $\delta z_l$ which propagates into theoretical
distance ratio calculations. In principle one should have accounted for them by suitable numerical simulations.
However, based on the experience gained on SNIa \citep{Perlmutter1999}, this effect is likely to be much smaller than  systematic errors discussed above.
Another straightforward solution based on
Poissonian statistics suggests that a sample size of order of a few
hundred lenses might reduce the line-of-sight 'noise' contamination
down to a few percent \citep{Kubo10}. However, our $D_{ds}/D_s$ data
set is really small, and its range of redshift is also limited.
Fortunately, with the ongoing of various systematic gravitational
surveys and more giant arc survey projects carried out by the
International X-ray Observatory (IXO) \citep{2010IXO}, extended
Roentgen Survey with an Imaging Telescope Array (eRosita)
\citep{Predehl10} and the Wide Field X-ray Telescope (WFXT)
\citep{2010WFXT} being under way,
the sample of strong lenses is growing rapidly, which may ease the
problem of line-of-sight contamination. Future observations will
definitely enlarge our set and make the method applied in this paper
more powerful.

\section*{Acknowledgments}
This work was supported by the National Natural Science Foundation
of China under the Distinguished Young Scholar Grant 10825313 and
Grant 11073005, the Ministry of Science and Technology national
basic science Program (Project 973) under Grant No. 2012CB821804, the
Fundamental Research Funds for the Central Universities and
Scientific Research Foundation of Beijing Normal University, and the
Polish Ministry of Science Grant No. N N203 390034.


\begin{thebibliography}{99}

\bibitem[Adelman-McCarthy et al.(2007)]{Adelman07} Adelman-McCarthy, J., et al. 2007, ApJS, 172, 634
\bibitem[Adelman-McCarthy et al.(2008)]{Adelman08} Adelman-McCarthy, J., et al. 2008, ApJS, 175, 297
\bibitem[Allam et al.(2004)]{Allam04} Allam, S. S., et al. 2004, AJ, 127, 1883
\bibitem[Allam et al.(2007)]{Allam07} Allam, S. S., et al. 2007, ApJ, 662, L51
\bibitem[Allen et~al.(2008)]{Allen08} Allen, S. W., et al. 2008, MNRAS, 383, 879
\bibitem[Amanullah et al.(2010)]{Amanullah} Amanullah, R., et al. 2010, ApJ, 716, 712 [arXiv:1004.1711]
\bibitem[Amati et~al.(2008)]{Amati2008} Amati, L., et~al. 2008, MNRAS, 391, 577
\bibitem[Bar-Kana(1996)]{Bar-Kana96} Bar-Kana, R. 1996, ApJ, 468, 17
\bibitem[Bassett \& Kunz(2004)]{Bassett04} Bassett, B. A., \& Kunz, M. 2004, PRD, 69, 101305
\bibitem[Biesiada, Godlowski \& Szydlowski(2005)]{Biesiada05} Biesiada, M.,Godlowski, W., \& Szydlowski, M. 2005, ApJ, 622, 28
\bibitem[Biesiada(2006)]{Biesiada06} Biesiada, M. 2006, PRD, 73, 023006
\bibitem[Biesiada et al.(2010)]{Biesiada10} Biesiada, M., Pi\'{o}rkowska, A., \& Malec, B. 2010, MNRAS, 406, 1055
\bibitem[Biesiada, Malec \& Pi\'{o}rkowska(2011)]{Biesiada11} Biesiada, M., Malec, B., \& Pi\'{o}rkowska, A. 2011, RAA, in print
\bibitem[Bonamente et~al.(2006)]{Bonamente2006} Bonamente et~al. 2006, ApJ, 647, 25
\bibitem[Borgani et~al.(1999)]{Borgani1999} Borgani, S., Rosati, P., Tozzi, P., \& Norman, C. 1999, ApJ, 517, 40
\bibitem[Breimer \& Sanders(1992)]{Breimer1992} Breimer, T.~G., \& Sanders, R.~H. 1992, MNRAS, 257, 97
\bibitem[Browne et al.(2003)]{Bro03} Browne, I.~W.~A., et al. 2003, MNRAS, 341, 13
\bibitem[Burles et~al.(2001)]{Burles2001} Burles, S., Nollett, K.~M., \& Turner, M.~S. 2001, ApJL, 552, L1
\bibitem[Caldwell et~al.(1998)]{Caldwell1998} Caldwell, R.~R., Dave, R., \& Steinhardt, P.~J. 1998, PRL, 80, 1582
\bibitem[Caldwell(2002)]{Caldwell2002} Caldwell, R.~R. 2002, PLB, 545, 23
\bibitem[Cao \& Zhu(2011b)]{Cao11b} Cao, S., \& Zhu, Z.-H. 2011, China Series G, 54, 12 [arXiv:1102.2750]
\bibitem[Cao \& Zhu(2011a)]{Cao11a} Cao, S., \& Zhu, Z.-H. 2011, A\&A, in press, arXiv:1105.6182
\bibitem[Cavaliere \& Fusco-Femiano(1976)]{Cavaliere1976} Cavaliere, A., \& Fusco-Femiano, R. 1976, A\&A, 49, 137
\bibitem[Chae(2003)]{Cha03} Chae, K.-H. 2003, MNRAS, 346, 746
\bibitem[Chae \& Mao(2003)]{CM03} Chae, K.-H., \& Mao, S. D. 2003, ApJ, 599, L61
\bibitem[Chae et al.(2004)]{Cha04} Chae, K.-H., Chen, G., Ratra, B., \& Lee, D.-W. 2004, ApJ, 607, L71
\bibitem[Chevalier \& Polarski(2001)]{Chevalier01} Chevalier, M., \& Polarski, D. 2001, IJMPD, 10, 213
\bibitem[Christlein(2000)]{Christlein00} Christlein, D.\ 2000, ApJ, 545, 145
\bibitem[Covone et~al.(2005)]{Covone05} Covone, G., et al. 2005, Submitted to A\&A [arXiv:0511332]
\bibitem[Dalal et al.(2005)]{Dalal05} Dalal, N., Hennavi, J. F., \& Bode, P. 2005, ApJ, 622, 99
\bibitem[Daly et~al.(2009)]{Daly2009} Daly, R.~A., et~al. 2009, ApJ, 691, 1058
\bibitem[Davis et~al.(2007)]{Davis2007} Davis, T.~M. et~al. 2007, ApJ, 666, 716
\bibitem[Di Pietro \& Claeskens(2003)]{Pietro03} Di Pietro, E., \& Claeskens, J. F. 2003, MNRAS, 341, 1299
\bibitem[Dvali et~al.(2000)]{DGP2000} Dvali, G., Gabadadze, G., \& Porrati, M. 2000, PLB, 485, 208
\bibitem[Eisenstein et al.(2005)]{Eisenstein05} Eisenstein, D. J., et al. 2005, ApJ, 633, 560
\bibitem[Fassnacht \& Cohen(1998)]{Fassnacht98} Fassnacht, \& Cohen, 1998, MNRAS, submitted
\bibitem[Feng et~al.(2005)]{Feng2005} Feng, B., Wang, X., \& Zhang, X. 2005, PLB, 607, 35
\bibitem[Freese \& Lewis(2002)]{Freese2002} Freese, K., \& Lewis, M. 2002, PLB, 540, 1
\bibitem[Gladders et~al.(2003)]{Gladders2003} Gladders, M.~D., et al. 2003, ApJ, 593, 48
\bibitem[Grillo et al.(2008)]{Grillo08} Grillo, C., Lombardi, M., \& Bertin, G. 2008, A\&A, 477, 397
\bibitem[Hennawi et~al.(2008)]{Hennawi2008} Hennawi, J.~F., et~al. 2008, AJ, 135, 664
\bibitem[Hinshaw et al.(2009)]{Hinshaw09} Hinshaw, G., et al. 2009, ApJS, 180, 225
\bibitem[Holanda, Lima \& Ribeiro(2010)]{HLB10} Holanda, R. F. L., Lima, J. A. S., \& Ribeiro, M. B., 2010, ApJ, 722, L233
\bibitem[Itoh et al.(1998)]{Itoh98} Itoh, N., Kohyama, Y., \& Nozawa, S. 1998, ApJ, 502, 7
\bibitem[Jin et al.(2000)]{Jin00} Jin, K.-J., Zhang, Y.-Z., \& Zhu, Z.-H. 2000, PLA, 264,335
\bibitem[Jones et~al.(2005)]{Jones2005} Jones, M.~E., et~al. 2005, MNRAS, 357, 518
\bibitem[Kamenshchik et al.(2001)] {Kamenshchik01}Kamenshchik, A., Moschella, U., \& Pasquier, V. 2001, PLB, 511, 265
\bibitem[Keeton et al.(1997)]{Keeton97} Keeton, C. R., Kochanek, C. S., \& Seljak, U. 1997, ApJ, 482, 604
\bibitem[Keeton et al.(2000)]{Keeton00} Keeton, C.~R., Christlein, D., \& Zabludoff, A.~I. 2000, ApJ, 545, 129
\bibitem[Keeton(2001)]{Kee01} Keeton, C.~R. 2001, ApJ, 561, 46
\bibitem[King et al.(1997)]{Kin97} King, L. J., et al. 1997, MNRAS, 289, 450
\bibitem[Kochanek(1992)]{Kochanek92} Kochanek, C. S., 1992, ApJ, 384, 1
\bibitem[Kochanek(1996)]{Koc96} Kochanek, C.~S. 1996, ApJ, 466, 638
\bibitem[Kochanek \& White(2001)]{KW01} Kochanek, C. S., \& White, M. 2001, ApJ, 559, 531
\bibitem[Komatsu et al.(2009)]{Komatsu09} Komatsu, E., et al. [WMAP Collaboration], 2009, AJS, 180, 330
\bibitem[Koopmans \& Treu(2002)]{Koopmans02} Koopmans, L. V. E., \& Treu, T. 2002, ApJ, 568, L5
\bibitem[Koopmans \& Treu(2003)]{Koopmans03} Koopmans, L. V. E., \& Treu, T. 2003, ApJ, 583, 606
\bibitem[Koopmans et al.(2006)]{Koopmans06} Koopmans, L. V. E., et al. 2006, ApJ, 649, 599
\bibitem[Koopmans et al.(2009)]{Koopmans09} Koopmans, L. V. E., et al. 2009, ApJ, 703, L51
\bibitem[Kowalski et~al.(2008)]{Kowalski2008} Kowalski, M., et~al. 2008, ApJ, 686, 749
\bibitem[Kubo et al.(2010)]{Kubo10} Kubo, J. M., et al. 2010, accepted by ApJL, arXiv:1010.3037v2
\bibitem[Linder(2003)]{Linder03} Linder, E. V. 2003, PRD, 68, 083503
\bibitem[Linder \& Roberts(2008)]{Linder08} Linder, E. V., \& Roberts, G. 2008, J. Cosmol. Astropart. Phys., 0806, 004
\bibitem[Lynds \& Petrosian(1986)]{Lynds1986} Lynds, R., \& Petrosian, V. 1986, AAS, 18, 1014
\bibitem[Mao \& Schneider(1998)]{MS98} Mao, S. D., \& Schneider, P. 1998, MNRAS, 295, 587
\bibitem[Martel et al.(2002)]{Martel02} Martel, H., Premadi, P., \& Matzner, R. 2002, ApJ, 570, 17
\bibitem[Mitchell et al.(2005)]{Mit05} Mitchell, J.~L., Keeton, C.~R., Frieman, J.~A., \& Sheth, R.~K. 2005, ApJ, 622, 81
\bibitem[Momcheva et al.(2006)]{Momcheva06} Momcheva, I., Williams, K., Keeton, C. R., \& Zabludoff, A. 2006, ApJ, 641, 169
\bibitem[Murray \& WFXT Team(2010)]{2010WFXT} Murray, S.~S., \& WFXT Team. 2010, AAS, 41, 520
\bibitem[Narayan \& Bartelmann(1996)]{Narayan96}Narayan, R. \& Bartelmann, M. 1996, In: ``Lectures on Gravitational Lensing'', arXiv:9606001
\bibitem[Newton et~al.(2011)]{Newton11} Newton, E., et al. 2011, accepted by ApJ, arXiv:1104.2608
\bibitem[Nozawa et al.(1998)]{Nozawa98} Nozawa, S., Itoh, N., \& Kohyama, Y. 1998, ApJ, 508, 17
\bibitem[Nozawa et al.(2006)]{Nozawa06} Nozawa, S., Itoh, N., Suda, Y., \& Ohhata, Y. 2006, Nuovo Cimento, 121 B, 487
\bibitem[Ofek et al.(2003)]{Ofek03} Ofek, E. O., Rix, H.-W., \& Maoz, D. 2003, MNRAS, 343, 639
\bibitem[Ono et~al.(1999)]{Ono1999} Ono, T., Masai, K., \& Sasaki, S. 1999, PASJ, 51, 91
\bibitem[Ota \& Mitsuda(2004)]{Ota2004} Ota, N., \& Mitsuda, K. 2004, A\&A, 428, 757
\bibitem[Paczynski \& Gorski(1981)]{Paczynski1981} Paczynski, B., \& Gorski, K. 1981, ApJL, 248, L101
\bibitem[Perlmutter et~al.(1999)]{Perlmutter1999} Perlmutter, S., et~al. 1999, ApJ, 517, 565
\bibitem[Pi\'{o}rkowska, Biesiada \& Malec(2010)]{Biesiada11b} Pi{\'o}rkowska, A., Biesiada, M. \& Malec, B., 2011, Acta Phys. Polon. B, 42, 2297
\bibitem[Predehl et~al.(2010)]{Predehl10} Predehl, P. et~al. 2010, arXiv:1001.2502
\bibitem[Press \& Schechter(1974)]{Press1974} Press, W.~H., \& Schechter, P. 1974, ApJ, 187, 425
\bibitem[Reese et~al.(2002)]{Reese2002} Reese, E.~D., et al. 2002, ApJ, 581, 53
\bibitem[Richard et~al.(2007)]{Richard07} Richard, J., et al. 2007, ApJ, 662, 781; Abell 68
\bibitem[Riess et~al.(1998)]{Riess1998} Riess, A.~G., et~al. 1998, AJ, 116, 1009
\bibitem[Riess et~al.(2004)]{Riess2004} Riess, A.~G., et~al. 2004, ApJ, 607, 665
\bibitem[Rosati et~al.(2002)]{Rosati2002} Rosati, P., Borgani, S., \& Norman, C. 2002, ARA\&A, 40, 539
\bibitem[Rusin \& Kochanek(2005)]{RK05} Rusin, D., \& Kochanek, C.~S. 2005, ApJ, 623, 666
\bibitem[Samushia \& Ratra(2008)]{Samushia08} Samushia, L., \& Ratra, B. 2008, ApJ, 680, L1
\bibitem[Schmidt et~al.(2004)]{Schmidt2004} Schmidt, R.~W., Allen, S.~W., \& Fabian, A.~C. 2004, MNRAS, 352, 1413
\bibitem[Schneider et~al.(1992)]{Schneider1992} Schneider, P., Ehlers, J., \& Falco, E.~E. 1992, Gravitational Lenses
\bibitem[Sereno(2002)]{Sereno2002} Sereno, M. 2002, A\&A, 393, 757
\bibitem[Sereno \& Longo(2004)]{Sereno2004} Sereno, M., \& Longo, G. 2004, MNRAS, 354, 1255
\bibitem[Sheth et al.(2003)]{She03} Sheth, R.~K., et al. 2003, ApJ, 594, 225
\bibitem[Spergel et~al.(2007)]{Spergel2007} {Spergel}, D.~N. {et~al.} 2007, ApJS, 170, 377
\bibitem[Stoughton et al.(2002)]{Sto02} Stoughton, C., et al. 2002, AJ, 123, 485
\bibitem[Sunyaev \& Zeldovich(1972)]{Sunyaev1972} Sunyaev, R.~A., \& Zeldovich, Y.~B. 1972, Comments on Astrophysics and Space Physics, 4, 173
\bibitem[Treu \& Koopmans(2004)]{Treu04} Treu, T., \& Koopmans, L. V. E. 2004, ApJ, 611, 739
\bibitem[Treu et al.(2006a)]{Treu06a} Treu, T., et al. 2006a, ApJ, 640, 662
\bibitem[Treu et al.(2006b)]{Treu06b} Treu, T., et al. 2006b, ApJ, 650, 1219
\bibitem[Treu et al.(2009)]{Treu09}Treu, T., et al. 2009, ApJ, 690, 670
\bibitem[Uzan et al.(2004)]{Uzan04} Uzan, J. P., Aghanim, N., \& Mellier, Y. 2004, PRD, 70, 083533
\bibitem[Walsh et~al.(1979)]{1979Natur} Walsh, D., Carswell, R.~F., \& Weymann, R.~J. 1979, Nature, 279, 381
\bibitem[Weinberg(1989)]{Weinberg} Weinberg, S. 1989, Rev. Mod. Phys., 61, 1
\bibitem[White \& Davis(1996)]{White96} White, R.~E.~\& Davis, D.~S.\ 1996, American Astronomical Society Meeting, 28, 1323
\bibitem[White et~al.(2010)]{2010IXO} White, N.~E., et al. 2010, arXiv:1001.2843
\bibitem[Yu \& Zhu(2010)]{Yu10} Yu, H., \& Zhu, Z.-H., accepted by RAA, arXiv:1011.6060
\bibitem[Zhang \&Zhu(2006)]{Zhang2006} Zhang, H., \& Zhu, Z.-H. 2006, PRD, 73, 043518
\bibitem[Zhu \& Wu(1997)]{Zhu97} Zhu, Z.-H., \& Wu, X.-P. 1997, A\&A, 324, 483
\bibitem[Zhu(2000a)]{Zhu00a} Zhu, Z.-H. 2000, MPLA, 15, 1023
\bibitem[Zhu(2000b)]{Zhu00b} Zhu, Z.-H. 2000, IJMPD, 9, 591
\bibitem[Zhu(2004)]{Zhu2004} Zhu, Z.-H. 2004, A\&A, 423, 421
\bibitem[Zhu \& Fujimoto(2004)]{2004Zhu2} Zhu, Z.-H., \& Fujimoto, M. 2004, ApJ, 602, 12
\bibitem[Zhu et~al.(2004)]{Zhu2004b} Zhu, Z.-H., Fujimoto, M., \& He, X. 2004, ApJ, 603, 365
\bibitem[Zhu \& Mauro(2008a)]{Zhu08a} Zhu, Z.-H., \& Mauro, S. 2008, A\&A, 487, 831
\bibitem[Zhu et al.(2008b)]{Zhu08b} Zhu, Z.-H., et al. 2008, A\&A, 483, 15


\end{thebibliography}
\end{document}